\documentclass[useAMS,usenatbib,referee]{biom}
\pdfoutput=1

\usepackage{amsmath,amssymb}
\usepackage{algorithm} 
\usepackage{algpseudocode} 
\usepackage{graphicx} 
\usepackage{xcolor}
\usepackage{multirow}
\usepackage[nodisplayskipstretch]{setspace}
\usepackage{titlesec}
\titlespacing{\section}{0pt}{2ex}{1ex}
\titlespacing{\subsection}{0pt}{1ex}{0ex}
\titlespacing{\subsubsection}{0pt}{0.5ex}{0ex}

\usepackage{enumitem}
\setlist{nolistsep}

\usepackage{epsfig}
\usepackage[normalem]{ulem}

\newcommand{\cm}[1]{\ignorespaces}

\def\bfw{\mathbf w}

\def\bfI{\mathbf I}

\def\bfW{\mathbf W}

\def\bfbeta{\boldsymbol \beta}

\def\bfSigma{\boldsymbol\Sigma}

\def\mE{\mathrm{E}}

\def\mE{\mathrm{E}}

\def\nde{\mathrm{NDE}}

\def\nie{\mathrm{NIE}}

\def\bSig\mathbf{\Sigma}
\makeatletter
\def\prop@space@setup{\prop@preskip=0pt\prop@postskip=0pt}
\makeatother
\newtheorem{prop}{Proposition}

\title[Mediation Analysis with Mixed Data]{A Class of Directed Acyclic Graphs with Mixed Data Types in Mediation Analysis}

\author
{Wei Hao\emailx{weihao@umich.edu} \\
	Department of Biostatistics, University of Michigan, Ann Arbor, Michigan, U.S.A.
	\and
	Canyi Chen\email{canyic@umich.edu} \\
	Department of Biostatistics, University of Michigan, Ann Arbor, Michigan, U.S.A.
	\and
	Peter X.-K. Song\emailx{pxsong@umich.edu} \\
	Department of Biostatistics, University of Michigan, 
	Ann Arbor, Michigan, U.S.A}

\graphicspath{{figs/}}
\begin{document}
	
	
	\date{{\it Received October} 2007. {\it Revised February} 2008.  {\it
			Accepted March} 2008.}
	
	
	
	\pagerange{\pageref{firstpage}--\pageref{lastpage}} 
	\volume{64}
	\pubyear{2008}
	\artmonth{December}
	
	
	\doi{10.1111/j.1541-0420.2005.00454.x}
	
	
	\label{firstpage}
	
	
	\begin{abstract}
		We propose a unified class of generalized structural equation models (GSEMs) with data of mixed types in mediation analysis, including continuous, categorical, and count variables. Such models extend substantially the classical linear structural equation model to accommodate many data types arising from the application of mediation analysis.  
		Invoking the hierarchical modeling approach, we specify GSEMs by a copula joint distribution of outcome variable, mediator and exposure variable, in which marginal distributions are built upon generalized linear models (GLMs) with confounding factors. We discuss the identifiability conditions for the causal mediation effects in the counterfactual paradigm as well as the issue of mediation leakage, and develop an asymptotically efficient profile maximum likelihood estimation and inference for two key mediation estimands, natural direct effect and natural indirect effect, in different scenarios of mixed data types. 
		The proposed new methodology is illustrated by a motivating epidemiological study that aims to investigate whether the tempo of reaching infancy BMI peak (delay or on time), an important early life growth milestone, mediate the association between prenatal exposure to phthalates and pubertal health outcomes. 
	\end{abstract}
	
	%
	
	\begin{keywords}
		Acyclicity; Causality; confounding; Copula; Mediation leakage; Sensitivity Analysis.   
	\end{keywords}
	
	
	\maketitle
	
	
	%
	
	\section{Introduction}
	\label{intro}
	In many biomedical studies, mediation analysis is the choice of method to investigate a putative mechanistic  pathway that involves two or more hypothesized causal factors. A mechanistic pathway under investigation is often pictorially presented by a directed acyclic graph (DAG), as shown in Figure \ref{fig: tri}(a), in which the acyclicity explains how the influence of exposure $X$ on outcome $Y$ is mediated through an intermediate variable $M$ or {\em a mediator}.  According to the acyclicity topology in a DAG, there are two paths from exposure $X$ to outcome $Y$: a direct path,  $X \to Y$,  and an indirect path via mediator $M$, $X \to M \to Y$, and the latter is called \emph{mediation pathway}. The topology of a DAG is modeled by three parameters $\alpha$, $\beta$ and $\gamma$  as illustrated in Figure \ref{fig: tri}(a). 
	The classical linear structural equation modeling approach under the multivariate normality \citep{baron1986moderator}, as demonstrated in  {Web Appendix A} under the DAG in Figure \ref{fig: tri}(b), adjusting confounding factors takes place only at univariate marginal distributions, while the path parameters $\alpha$, $\beta$ and $\gamma$ exclusively characterize the acyclicity topology via the covariance of the joint distribution.  Such insights navigate our extensions considered in this paper so that the resulting methodology can continue to enjoy technical ease in statistical analysis as well as desirable interpretability of mediation effects. Unfortunately, most of existing extensions to the nonlinear non-normal models in the literature have compromised such desirable modeling features, leading to daunting technical challenges and much confusion in interpreting model parameters. This paper aims to address this significant gap in the current arsenal of mediation analysis. 
	
	The primary objective of a mediation analysis is twofold: (i) to evaluate the effect of exposure on outcome when the mediator is held constant, i.e. the path $X \to Y$, termed as {\em direct effect} (DE); and (ii) to evaluate the effect of exposure on outcome through the mediator, i.e. the path $X \to M \to Y$, termed as {\em mediation effect or indirect effect} (IE). These estimands are originally proposed by \cite{baron1986moderator} through a system of linear regression models, known as the structural equation modeling (SEM), which have been extensively discussed and routinely applied in practice. An important extension of the SEM merged recently within the counterfactual outcome framework \citep{robins1992identifiability,pearl2001direct}, which has attracted considerable attention to the development of causal mediation inference.  In the causality context, both DE and IE are re-established as the so-called natural direct effect (NDE) and natural indirect effect (NIE) under some identifiability conditions, including the sequential ignorability assumption; for example, \cite{imai2010identification, coffman2012assessing, preacher2015advances}. 
	
	\begin{center}
		\vspace{-0.3cm}
		{ [Figure 1 about here.]}
		\vspace{-0.3cm} 
		\begin{figure}
			\centering
			\includegraphics[width=4.5in]{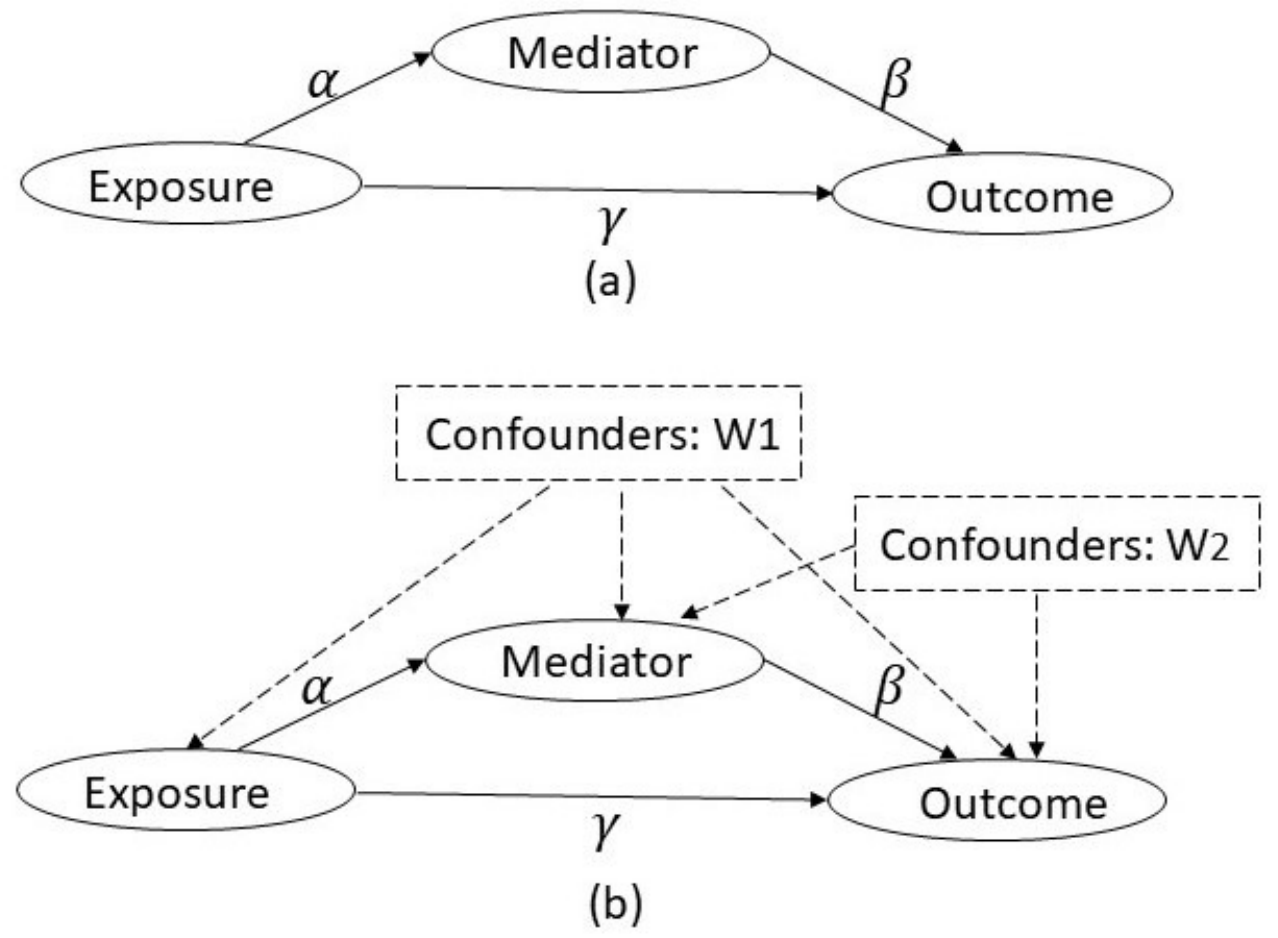}
			\caption{Directed acyclic graphs involving exposure $X$, mediator $M$ and outcome $Y$: (a) without confounders, and (b) with confounders.}
			\label{fig: tri}
		\end{figure}
	\end{center}

	Motivated from an environmental epidemiological study with a binary mediator in Section~\ref{real data}, we propose an extension of SEM to accommodate non-normal mixed data types by a joint modeling approach based on a copula \citep{xue2000multivariate,song2009joint}.  Our extension gives rise to a full probability model that incorporates the DAG dependence topology, denoted by $\pi(x, m, y \mid \bfW)$, where $\pi(\cdot)$ is  a certain suitable joint probability density function to be specified later in the paper and $\bfW$ is a set of confounders. Our proposed methodology differs from  the existing approaches that focus merely on the conditional expectations of outcome and mediators such as $E(Y | X=x, M=m, \bfW)$ and $E(M| X=x, \bfW)$ in which no explicit joint probability distributions are specified in the mediation analysis.  One advantage of our approach pertains to its flexibility of analyzing different data types, accommodating continuous or discrete variables. 
	Performing mediation analysis of non-normal data outside the scope of multivariate normal distributions is not trivial because the underlying joint probability distribution for the data generation may not be uniquely defined, which subsequently affects the validity of statistical inference that relies on a legitimate probability measure to quantify confidence interval coverage or $p$-value.  In this non-normal data context, existing approaches that specify only the conditional expectations may suffer from significant ambiguity for the joint distribution that may  lead to an invalid statistical inference. 
	
	In the paper we attempt to overcome such a limitation by introducing an explicit joint distribution that can model the acyclicity of the mediation effects. Our approach defines a unified framework of multivariate non-normal distributions generated from the Gaussian copula, which is known to be marginally closed \citep{xue2000multivariate}. 
	Refer to the details in {Web Appendix E}. 
	There are some scattered works concerning {\it ad hoc} cases,  such as categorical mediator and/or outcome, or continuous but non-normal mediator and/or outcome  \citep{vanderweele2014mediation,vanderweele2010odds,albert2011generalized,huang2004statistical,tingley2014mediation}. 
	Usually, stronger model assumptions are imposed when data are of mixed types. Moreover, as becoming evident throughout this paper, causal interpretations appear to be model-specific.  
	When the joint normality is unavailable, there lacks of a unified framework  of mediation analysis to handle data of mixed types.
	
	Following the formulation of 
	the copula regression~\citep{song2009joint}, we propose a unified class of \emph{generalized structural equation models (GSEMs)} for non-normal data. 
	These GSEMs provide three new methodological advantages: first, this unified approach allows exposure, mediator and outcome to be different types, either categorical, discrete or continuous, in which model constructs and parameters are adaptive to various data types.  Second, our unified joint modeling approach permits the derivation of a coalescent joint likelihood function over different data scenarios, facilitating flexibility and rigorousness than existing \emph{ad hoc} conditional expectation based approaches.   Third, taking advantage of a unified specification, we consolidate the estimation and inference by Joe's asymptotically efficient two-stage estimation method~\citep{joe2005asymptotic}. 
	
	This paper is organized as follows.  Section \ref{real data} discusses the motivating ELEMENT study. Section \ref{model} introduces GSEMs. Section \ref{estimation} focuses on estimating model parameters and causal estimands under eight scenarios of mixed data types, followed by three examples in  Section \ref{three eg}. Sections \ref{simul} and \ref{data analysis} present simulation experiments and a mediation analysis of the motivating data. Section \ref{concluding} concerns some concluding remarks.  Detailed technical derivations are included in the Supplementary Material.
	
	\section{ELEMENT Cohort Study}\label{real data}
	Our new methodology development is motivated by the ``Early Life Exposures in Mexico to ENvironmental Toxicants" (ELEMENT) cohort study that consists of three birth cohorts of 1,643 mother-child pairs during pregnancy or delivery from maternity hospitals in Mexico city between 1993 and 2004 \citep{hu2006fetal, perng2019early}. One of objectives is to study how environmental toxicants, such as phthalates,  affect maternal and child health outcomes. Known as endocrine disrupting chemicals (EDCs), phthalates are a group of chemicals mostly used in plastics, and exposure to elevated levels of these chemicals during pregnancy has been associated with adverse health outcomes for both mothers and children \citep{qian2020endocrine}. A previous study \citep{zhou2021} has found that third trimester maternal phthalate exposures of MEHHP, MEOHP and MIBP are linked to delayed infancy Body Mass Index (BMI) peak. In another recent study, a delayed reach of infancy BMI peak has been shown to be associated with higher cardiometabolic risk during peripuberty \citep{perng2018associations}. Linking these two related findings forms a potential mechanistic pathway: prenatal phthalate exposure $\rightarrow$ growth delay in early life $\rightarrow$  cardiometabolic risk later in life, as shown in Figure \ref{co: fig: realdata}.  Thus, it is of great interest to investigate this hypothesized mechanistic pathway by means of mediation analysis, in which a technical challenge pertains to a binary mediator as the timing of children reaching their infancy BMI peak (on time or delayed).  Note that infancy BMI peak is an important milestone for the early-life growth associated with obesity development of children in later life \citep{jensen2015infant,bornhorst2017potential}.
	
	The data used to investigate the mediation effect of children growth marker contains 205 mother-child pairs with 97 boys and 108 girls, where 
	the mean ages of mothers at birth, and children at cardiometabolic risk measurements are 27.7 yrs and 10.1 yrs, respectively. The hypothesized  mediator, the timing of infancy BMI peak,  is measured by age (in month) at which a child reaches his/her BMI peak before 3 years old \citep{baek2019bayesian}. 
	A child who fails to reach such peak before 3 years old is deemed as delayed growth ($M=1$) or on time ($M=0$), and 27.8\% of children in the data exhibit the delayed reach of infancy BMI peaks.
	
	The exposure variables include urinary concentrations of three EDCs (MEHHP, MEOHP, and MIBP) measured, respectively, at the second trimester among 177 women and third trimester among 202 women. These three EDCs at the third trimester were previously found to be positively associated with delayed infancy BMI peak \citep{zhou2021}. MetS z-score is the outcome of interest, which is a composite cardiometabolic risk biomarker measured on children during peripuberty; it is formed as an
	average of five z-scores from waist circumference, fasting glucose, C-peptide, ratio of triglycerides and high-density lipoprotein cholesterol and average of systolic and diastolic blood pressures 
	\citep{perng2018associations}.
	
	Figure \ref{co: fig: realdata} shows a mechanistic pathway: continuous phthalate concentration $X$ $\rightarrow$ binary growth marker $M$ $\rightarrow$ continuous cardiometabolic outcome $Y$, along with two sets of confounders $\bfW_1$ and $\bfW_2$. Vector $\bfW_1$ includes maternal age at delivery, education, and marital status; these mother's intrinsic characteristics may influence prenatal exposures, growth mediator and cardiometabolic health outcomes. Vector $\bfW_2$ include breastfeeding duration, parity, child sex, gestational age at delivery, and birth weight that occur after birth, which may influence children's growth mediator and their health outcomes. Differentiating  pre- and post-exposure covariates is essential to adjust for confounding in assessing mediation effects.
	\begin{center}
		{[Figure 2 about here.]}
		\begin{figure}
			\centering
			\includegraphics[width=6.5in]{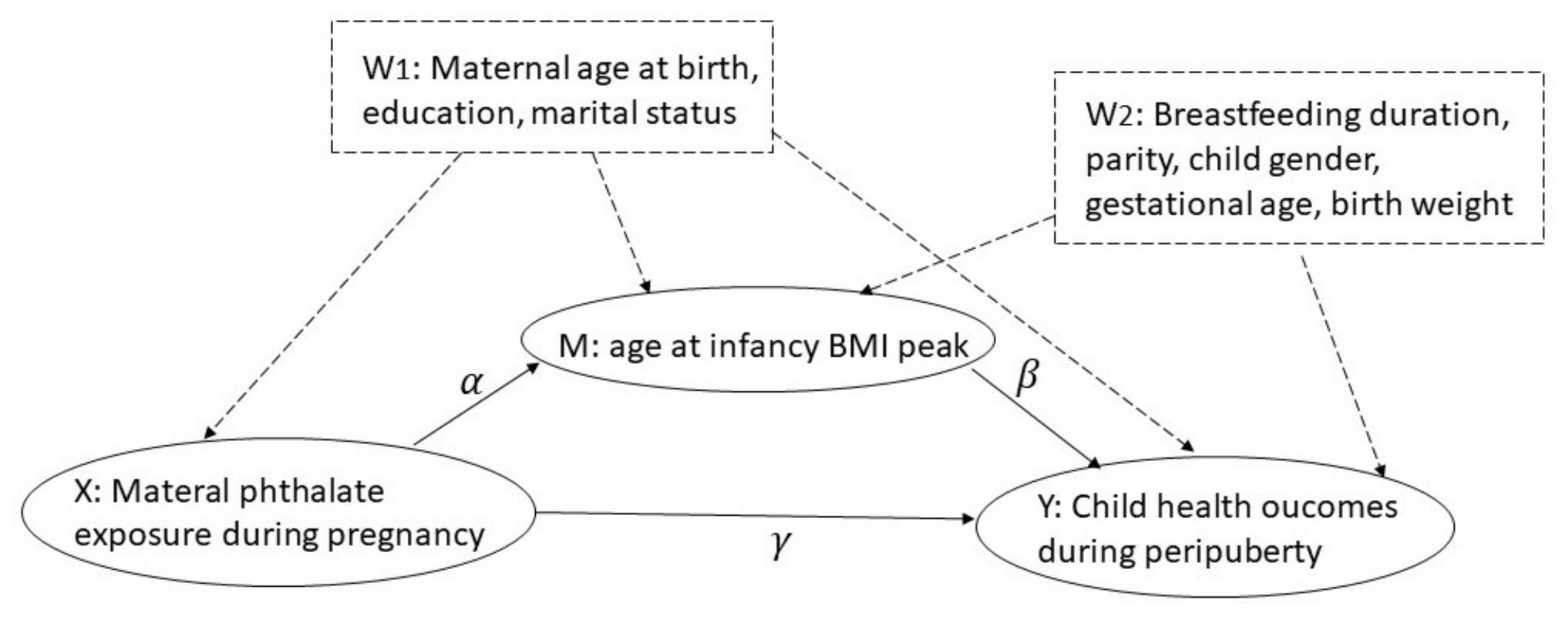}
			\caption{Association between phthalate exposures and health outcomes mediated by whether child's timing of reaching BMI peak is delayed or not.}
			\label{co: fig: realdata}
		\end{figure}
	\end{center}
	\section{Generalized Structural Equation Models}
	\label{model}
	
	\subsection{Framework}\label{model: framework} 
	\noindent  We begin with a discussion on our general modeling approach to constructing a joint distribution for three variables $(X, M, Y)$ of mixed types under the graphic topology of acyclic direct graph (DAG) shown in Figure~\ref{fig: tri}(a).  Such joint distribution is specified by a hierarchical modeling approach, which enables us to define a class of generalized structural equation models (GSEMs) to address practical needs in mediation analyses with data of mixed types. The key feature in the classical linear normal SEM is that the covariance matrix of $(X,M,Y)$, denoted by, $\Gamma$, takes a special form to satisfy the acyclicity topology of a DAG in the classical SEM as discussed in  {Web Appendix A}. That is, the acyclicity of the DAG is described by a lower-triangular matrix $\Theta$ that gives rise to a covariance matrix as follows: 
	\begin{equation}
		\label{eq: cov} 
		\Gamma = \left(\bfI - \Theta\right)^{-1} \bfSigma \left(\bfI - \Theta\right)^{-\top},~\Theta = \left(\begin{array}{ccc}
			0 & 0 & 0\\
			\alpha & 0 & 0\\
			\gamma & \beta & 0
		\end{array}\right),
	\end{equation}
	where $\bfI$ is the identity matrix, and $\bfSigma = \mbox{diag}(\sigma_x^2, \sigma_m^2, \sigma_y^2)$ is a diagonal matrix of marginal variances, each for one variable. The proposed extension will incorporate the above specification of $\Gamma$ in the formulation of a valid joint probability distribution for $X$, $M$, $Y$, resulting in a class of nonlinear non-normal SEMs, termed as \emph{generalized structural equation models} (GSEMs) in this paper. The formulation of GSEMs is expected to have following features:
	\begin{itemize}
		\item[(i)] The class of GSEMs is flexible with different types of marginal distributions so to accommodate continuous, discrete or categorical variables $X, M, Y$. 
		\item[(ii)] The class of GSEMs incorporates an explicit form of the covariance given in equation (\ref{eq: cov}). Such matrix form is deemed critical importance to model the acyclicity topology of a DAG and to interpret parameters pertinent to  mediation pathways. 
		\item[(iii)] Similar to the classical SEM discussed in  {Web Appendix A}, confounders do not enter the covariance matrix $\Gamma$; rather, they are only adjusted at marginal distributions of individual variables.  This consideration suggests a hierarchical model specification where the marginal and dependence parameters enter different hierarchies of GSEMs. 
		\item[(iv)] The classical linear normal SEM is a special case of the proposed GSEMs. 
	\end{itemize}
	
	These modeling characteristics above, as shown in the paper, can be incorporated simultaneously into the proposed joint distribution by the copula modeling approach. \cite{song2009joint} has demonstrated a similar approach to constructing a class of vector GLMs (VGLMs) with data of mixed of types.  Here, we consider a more complicated formulation than that of VGLMs in which the dependence matrix takes a restricted form given in equation (\ref{eq: cov}) and confounding factors are only allowed to enter the model in the hierarchies of marginal distributions. Indeed, the Fr{\'e}chet's theory of constructing multivariate distribution by bivariate distributions is the theoretical basis to ensure the validity of our proposed hierarchical models \citep{joe2014dependence}. 
	
	Copula models provide a natural hierarchical modeling framework to satisfy the above modeling requirements.  According to Sklar's Theorem \citep{sklar1959fonctions}, when $(X,M,Y)$ are continuous random variables, their joint distribution can be expressed as
	$F(x, m, y) = C\{F_x(x), F_m(m), F_y(y); \Gamma\}$,
	where $F_j(\cdot)$ and $f_j(\cdot)$ are the cumulative distribution function and density function, respectively, and $C(\cdot)$ is a suitable copula function that is independent of marginal parameters. This expression provides a hierarchical modeling framework, as desired.  To incorporate the covariance structure given in (\ref{eq: cov}) in the copula function, in this paper we choose the Gaussian copula \citep{xue2000multivariate},
	\begin{equation}\label{eq: Gaussian}
		C(u_1, u_2, u_3) = \Phi_3\{\Phi^{-1}(u_1),\Phi^{-1}(u_2),\Phi^{-1}(u_3); \Gamma\},
	\end{equation}
	where $\Phi_3(\cdot ; \Gamma)$ is a trivariate Gaussian distribution function with mean zero and correlation matrix $\Gamma$, and $\Phi(\cdot)$ is the standard univariate Gaussian distribution function and $\Phi^{-1}(\cdot)$ is the corresponding quantile function. Since all marginal parameters, including the variance parameters, are exclusively embedded in the marginal distributions, the dependence matrix $\Gamma$ contains only correlation parameters without any marginal variance parameters.  Parameters in $\Gamma$ represent rank-based correlations, more general dependencies than Pearson correlations. 
	
	The construction of GSEMs is based on a joint distribution generated by the Gaussian copula model (\ref{eq: Gaussian}) that accommodates either continuous or discrete marginal distributions as well as a special $\Gamma$ satisfying (\ref{eq: cov}). 
	The resulting model can be used to handle variables of mixed types in mediation analyses.  Below we consider two  settings separately: 
	GSEMs without confounders, and GSEMs with confounders in observational studies.

	\subsection{GSEMs under No Confounders}
	\label{model: douran} 
	\noindent   We first consider the most favorable scenario: there are no confounders involved in the relationships among exposure, mediator and outcome. One such case can be achieved by randomization of both $X$ and $M$ \citep{preacher2015advances}.  This is the simplest setting in the mediation analysis to introduce the GSEM and study related properties. Here we assume each marginal follows one generalized linear model (GLM); that is, the marginal distributions of $X$, $M$ and $Y$ are elements of the family of the exponential dispersion (ED)  models \citep{jorgensen1987exponential} denoted by,
	$X \sim ED(\mu_x,\phi_x)$,
	$M \sim ED(\mu_m,\phi_m)$, and
	$Y \sim ED(\mu_y,\phi_y)$,
	where $ (\mu_x,\mu_m,\mu_y)$, and $(\phi_x,\phi_m,\phi_y)$  are the respective mean parameters and dispersion parameters, all being free of covariates.  Denote by $F_x(x)$, $F_m(m)$ and $F_y(y)$ the cumulative  distribution functions of $X$, $M$ and $Y$, respectively. Setting $u_1 = F_x(X), u_2 = F_m(M), u_3 = F_y(Y)$, we can obtain a joint distribution $\pi(x,m,y)$ of $(X, M, Y)$ via the Gaussian copula function $C(u_1, u_2, u_3; \Gamma)$ in \eqref{eq: Gaussian}.  It can be shown that the resulting joint distribution under assumptions \eqref{eq: cov} and \eqref{eq: Gaussian} may be equivalently generated from a latent variable representation below in \eqref{eq: general_model} based on latent normal random variables $Z_x$, $Z_m$ and $Z_y$ satisfying:  
	\begin{eqnarray}
		Z_x \sim N(0, 1), ~
		Z_m\mid Z_x  \sim N(\alpha Z_x, 1),~
		Z_y \mid Z_x, Z_m  \sim N(\gamma Z_x + \beta Z_m, 1).\label{eq: general_model}
	\end{eqnarray}
	
	The correspondence between the observed variables $(X,M,Y)$ and the latent variables $(Z_x, Z_m, Z_y)$ is operated by the monotonic marginal quantile transformations: $X= F_x^{-1}\{\Phi(Z_x)\}$, $M= F_m^{-1}\{\Phi(Z^*_m)\}$ and $Y= F_y^{-1}\{\Phi(Z^*_y)\}$, where $Z_m^*=Z_m/\tau_m$, $Z_y^*=Z_y/\tau_y$, $\tau_m=\sqrt{\alpha^2+1}$, and $\tau_y=\sqrt{(\gamma+\alpha\beta)^2+\beta^2+1}$. 
	The marginal distributions of $Z_x$, $Z^*_m$ and $Z^*_y$ all follow the standard normal distribution, and the joint distribution of $Z_x$, $Z_m^*$ and $Z_y^*$ is a trivariate normal distribution; refer to {Web Appendix B} for the detailed derivation.   
	
	If $X$ is a continuous variable, $X= F_x^{-1}\{\Phi(Z_x)\}$ becomes $Z_x=\Phi^{-1}(F_x(X))$, namely the quantile of the standard normal distribution. If $X$ is a discrete variable with non-zero point probability mass at $0,1,2,\cdots$, then $X=\sum_{x=0}^{\infty}xI(F_x(x-1)\leq\Phi(Z_x)<F_x(x))$. This implies that event $\{X=x\}$ is equivalent to event $\{\Phi^{-1}(F_x(x-1))\leq Z_x<\Phi^{-1}(F_x(x))\}$ where
	for simplicity, the corresponding lower and upper limits are denoted by $\Phi^{-1}(F_x(x-1))=l_x$, and $\Phi^{-1}(F_x(x))=u_x$ in the rest of this paper. Similar arguments and notations are applied to the other two variables $M, Y$ under latent variables $Z_m^*$ and $Z_y^*$, respectively. For example, $l_m$ and $u_m$ are the lower and upper quantile limits in the case of mediator $M$, and so on.  Since $X$, $M$ and $Y$ may be either continuous or discrete, there are a total of eight scenarios. For instance, ``CCC", ``CDC" and ``CCD" represent, respectively, (i) $X$, $M$, and $Y$ are continuous; (ii) $X$ and $Y$ are continuous while $M$ is discrete; and (iii) $X$ and $M$ are continuous while $Y$ is discrete. {Web Table 1} presents the expressions of the corresponding likelihood functions in all scenarios. See the detailed derivation of $\pi(X, M, Y)$ in {Web Appendix E}.

	Now we derive natural direct effect (NDE) and natural indirect effect (NIE) under the above GSEM. We adopt the potential outcome framework from the causal inference literature \citep{splawa1990application,rosenbaum1983central,pearl2001direct,robins1992identifiability}. Following the counterfactual notions \citep{vanderweele2009conceptual}, let $Y(x,m)$ denote the counterfactual outcome that would have been observed for a subject had the exposure $X$ been set to value $x$ and mediator $M$ to the value $m$. Let $M(x)$ be the counterfactual value of mediator had the exposure $X$ been set to $x$. Then, $\mE\{Y(x_a,M(x_b))\}$ is the expected outcome of $Y$ had the exposure been set to $x_a$ and mediator been set to $M(x_b)$,  namely $\mE\{Y(x_a,M(x_b))\}=\mE_M\{\mE_Y(Y|M,X=x_a)|X = x_b\}.$ Both estimands NDE and NIE for a change of $X$ from $x_0$ to $x_1$ are given by, 
	$\nde(x_0,x_1) =\mE\{Y(x_1,M(x_0))\}- \mE\{Y(x_0,M(x_0))\}$, and 
	$\nie(x_0,x_1) = \mE\{Y(x_1,M(x_1))\}- \mE\{Y(x_1,M(x_0))\}$. As seen in Section~\ref{three eg}, these estimands are  determined by the DAG parameters given a specific GSEM.  
	
	\subsection{GSEMs in Observational Studies}
	\label{model: obs} 
	\noindent  In the setting of non-experimental design, such as observational studies,  confounding factors are used to adjust for confounding bias. Let $\bfW= (\bfW^{\top}_1,\bfW^{\top}_2)^{\top}$ denote a vector of all the confounding factors available in a dataset, where $\bfW_1 = (W_{1,1},\ldots, W_{1,p_1})^{\top}$ influences $X$, $M$ and $Y$, and $\bfW_2 = (W_{2,1},\ldots, W_{2,p_2})^{\top}$ only influences $M$ and $Y$; see Figure~\ref{fig: tri}(b).  For simplicity, the first element of $\bfW$ is set to 1 for the intercept. For the identification of NDE and NIE, we adopt the same assumptions in the literature \citep{vanderweele2009conceptual,VanderWeele:2015}: (i) there are no unmeasured confounders of the associations between $X$ and $M$, between $M$ and $Y$, and between $X$ and $Y$, respectively; (ii) no confounders of $M$ and $Y$ that are influenced by $X$; or equivalently the sequential ignorability assumption. The identifications of NDE and NIE under sequential ignorability can be established using the same techniques discussed in \cite{imai2010identification},  and details can be found in {Web Appendix C}. For the sensitivity analysis of sequential ignorability, we follow the work in \cite{imai2010identification}, and derive the covariance matrix under the latent variable representation with $Z_x$, $Z_m^*$ and $Z_y^*$ when the sequential ignorability assumption is violated.  The related details can be found in {Web Appendix D}. As discussed above, we allow confounders to enter the univariate GLMs for the marginals of the joint distribution of the following form: 
	$X \mid \bfW_1 \sim ED(\mu_x,\phi_x)~\mbox{with}~ g_x(\mu_x)=\bfW_1^\top\bfbeta_x$,  
	$M\mid \bfW \sim ED(\mu_m,\phi_m)~\mbox{with}~g_m(\mu_m)=\bfW^\top\bfbeta_m$, and 
	$Y \mid \bfW \sim ED(\mu_y,\phi_y)~\mbox{with}~g_y(\mu_y) =\bfW^\top\bfbeta_y$, 
	where $\bfbeta_x, \bfbeta_m$ and $\bfbeta_y$ are vectors of regression coefficients, and $g_x$, $g_m$ and $g_y$ are the respective link functions. The corresponding latent variable representation can be established with little effort. Moreover, both NDE and NIE are extended to the forms of conditional NDE and NIE given $\bfW=\bfw$ given as follows:
	\begin{align}
		\nde(x_0,x_1; \bfw) &=\mE\{Y(x_1,M(x_0))\mid \bfW=\bfw\}- \mE\{Y(x_0,M(x_0))\mid \bfW=\bfw\},\nonumber\\
		\nie(x_0,x_1; \bfw)& = \mE\{Y(x_1,M(x_1))\mid \bfW=\bfw\}- \mE\{Y(x_1,M(x_0))\mid \bfW=\bfw\}.\label{eq: cond eff def}
	\end{align}
	It is worth reiterating that in the proposal GSEMs, 
	covariates are used to adjust individual nodes marginally not edges of the DAG in Figure \ref{fig: tri}(b), and the latter concerns the acyclicity or the second moments of the joint distribution.  Here the GLM modeling of exposure $X$  is optional; scientifically, it is often the case in an observational study that covariates may affect exposure $X$, which, however, has been largely ignored in the existing literature. 
	
	\section{Estimation} \label{estimation}    
	\subsection{ Estimation of Model Parameters}\label{estimation: para}
	\noindent  We focus on a GSEM with covariates in the marginal GLMs discussed in Section \ref{model: obs}. Consider a dataset of $n$ $i.i.d.$ observations, $(X_i, M_i, Y_i, \bfW_i), i=1, \cdots, n.$ The maximum likelihood estimation (MLE) would be the method of choice given that the exact likelihood function is available. To account for computational efficiency to enhance practical usefulness of the proposed GSEM, we adopt a profile MLE of asymptotic efficiency studied by \cite{joe2005asymptotic}, among others,  which is well known in the copula literature.  Following the methodology of inference function for margins (IFM) \citep{xu1996statistical,joe2005asymptotic,shih1995inferences,ferreira2014modified,ko2019model}, we implement an asymptotically efficient two-stage profile likelihood  estimation procedure for parameter estimation in the GSEM. 
	
	The first stage of IFM runs a univariate GLM of $X_i$ marginally on covariates $\bfW_{1,i}$ and returns estimates $\hat\bfbeta_x$ and $\hat\phi_{x}$, as well as the individual fitted values $\hat\mu_{x_i}$. 
	Then we calculate individual $Z_{x_i}=\Phi^{-1}(\hat F_x(X_i|\bfW_{1i}))$ where the estimated CDF $\hat F_x(x|\bfw)$ is obtained by the ED distribution with both estimated mean $\hat\mu_{x_i}$ and dispersion $\hat\phi_x$. $Z_{x_i}$ could be either a unique value if variable $X$ is continuous or a range if it is discrete with lower and upper limits. Repeat the same marginal GLM analysis for the other two variables $M$ and $Y$ to obtain the $Z^*_{m_i}$ and $Z^*_{y_i}$.  Note that the GLM analysis on variable $X$ is optional when the conditional GSEM is specified under conditioning on $X$. 
	
	In the second stage of IFM, $Z_{x_i}$, $Z_{m_i}^*$ and $Z_{y_i}^*$ are plugged into the likelihood of the GSEM to create a profile log-likelihood function, $\sum_{i=1}^{n}\log\pi(X_i, M_i, Y_i)$, that is a function of  a function of three DAG parameters $\alpha$,  $\beta$ and $\gamma$. See the details in {Web Table 1}.  We call the R function \texttt{optim} to search for the estimates $\hat\alpha$, $\hat\beta$ and $\hat\gamma$ that minimize the negative profile log-likelihood function. In the implementation, we choose ``Nelder-Mead" algorithm  \citep{nelder1965simplex}, which only uses function values in the optimization.  The initial values used to start the search are set at 0 for all three parameters.  
	
	\subsection{Estimation of Mediation Causal Estimands}\label{estimation: eff}
	
	\noindent  To estimate the conditional NDE and NIE, we can calculate $\mE\{Y(x_a,M(x_b)\}$  for given $x_a$ and $x_b$ under each of the eight scenarios listed in {Web Table 1}. The detailed derivations are available in {Web Appendix F}. Notably, these mediation causal estimands are calculated for a representative individual with mean covariate values for continuous covariates, and/or a stratum for categorical covariates. In some cases, we can obtain the closed-form expressions for the expectations.  In those cases where the closed-form expressions are unavailable, we invoke numerical integral techniques based on the ``Monte Carlo" method to approximate the integrals in the calculations of NDE and NIE.
	
	\subsection{  Bootstrap for  Confidence Interval}\label{estimation: CI}
	\noindent As suggested in the literature, jackknife or bootstrap methods are recommended for statistical inference in the invocation of the asymptotically efficient profile likelihood estimation above. Here the bootstrap method is opted because the expectations in both conditional NDE and NIE appear so complex that it is hard to analytically derive the standard errors of these estimands using the Delta method. In the implementation, we consider both parametric and nonparametric bootstrap approaches   \citep{efron1987better} to obtain the 95\% confidence intervals (CI) for the DAG parameters $\alpha$, $\beta$ and $\gamma$ as well as the mediation causal estimands, NDE and NIE. In our empirical studies, we generate 500 bootstrap samples, each producing the estimates of $\alpha$, $\beta$ and $\gamma$ as well as NDE and NIE. Summarizing these 500 estimates, we construct a 95\% CI for one parameter by the 2.5 percentiles and 97.5 percentiles. In the use of parametric bootstrap, 500 bootstrap datasets are generated from the fitted GSEM under the estimates $\hat\alpha$, $\hat\gamma$ and $\hat\gamma$. In the nonparametric bootstrap approach, the bootstrap datasets are generated from randomly drawing of the observations with replacement.
	
	\section{Three Examples}\label{three eg}
	\noindent To illustrate the proposed method, we present three examples out of eight scenarios with mixed data types. CCC: $X$, $M$ and $Y$ are all continuous normal variables. CDC: $X$ and $Y$ are continuous normal variables while $M$ is a binary Bernoulli variable. CCD: $X$ and $M$ are continuous normal variables while Y is a binary Bernoulli variable. The proposed GSEM enables us to obtain the closed analytic forms of NDE and NIE in the two cases of CCC and CDC. In the CCD case, with the closed forms of NDE and NIE being unavailable we use the numerical integration method in the calculation. 
	
	\begin{prop}\label{prop: ccc}
		Suppose that $X$ and $M$ and $Y$ given $\bfW$ are normally distributed with marginal means $\mu_x = \bfW_1^\top \bfbeta_x, \mu_m = \bfW^\top \bfbeta_m,\mu_y = \bfW^\top \bfbeta_y$ and marginal variances $\sigma_x^2, \sigma_m^2, \sigma_y^2$. Then mediation causal estimands for a change of $X$ from $x_0$ to $x_1$ are given as follows:  
		\begin{align}
			\nde(x_0,x_1; \bfw) =\frac{\sigma_y\gamma(z_{x_1}-z_{x_0})}{\tau_y},~\mbox{and}~
			\nie(x_0,x_1; \bfw)=\frac{\sigma_y\alpha\beta(z_{x_1}-z_{x_0})}{\tau_y},\nonumber
		\end{align}
		where $z_x=(x-{\bfW_1^\top\bfbeta_x})/\sigma_x$ and $\tau_y=\sqrt{(\gamma+\alpha\beta)^2+\beta^2+1}$. Consequently, $ \nde(x_0,x_1; \bfw)=0$ if and only if $\gamma=0$; $ \nie(x_0,x_1; \bfw)=0$ if and only if $\alpha=0$ or $\beta=0$. 
	\end{prop}
	These equivalencies coincide with the properties known for the classical SEM in that covariates  $\bfW$ do not drive the acyclicity of DAG nor the estimands, NDE and NIE.

	\begin{prop}\label{prop: cdc}
		Suppose that $X$ and $Y$ given $\bfW$ are normally distributed with marginal means $\mu_x = \bfW_1^\top \bfbeta_x, \mu_y = \bfW^\top \bfbeta_y$ and marginal variances $\sigma_x^2, \sigma_y^2$, and that binary $M$ given $\bfW$ is Bernoulli distributed with the probability of success satisfying a logistic model of the form, $\mbox{expit}(\bfW^\top \bfbeta_m)$.  Then the mediation causal estimands a change of $X$ from $x_0$ to $x_1$ are given by 
		\begin{align}
			\nde(x_0,x_1; \bfw)&=\frac{\sigma_y(\gamma+\alpha\beta)(z_{x_1}-z_{x_0})}{\tau_y}-\frac{\sigma_y\beta}{\tau_y}\left(\frac{p_{z_{x_0}}}{p_{z_{x_1}}}-\frac{1-p_{z_{x_0}}}{1-p_{z_{x_1}}}\right)d_{z_{x_1}}(\bfw),\nonumber\\
			\nie(x_0,x_1; \bfw)&=\frac{\sigma_y\beta}{\tau_y}\left(\frac{p_{z_{x_0}}}{p_{z_{x_1}}}-\frac{1-p_{z_{x_0}}}{1-p_{z_{x_1}}}\right)d_{z_{x_1}}(\bfw),\nonumber
		\end{align}
		where $u_m(\bfw)=\Phi^{-1}(F_m(0))=-\Phi^{-1}\left(\mbox{expit}({\bfw^\top\bfbeta_m})\right)$, $p_x(\bfw)=\Phi(\tau_mu_m(\bfw)-\alpha x)$, $d_x(\bfw)=\phi(\tau_mu_m(\bfw)-\alpha x)$. Consequently,  $\nde(x_0,x_1; \bfw)=0$ if $\gamma=\beta=0$ or $\gamma=\alpha=0$, and   $\nie(x_0,x_1; \bfw)=0$ if $\alpha=0$ or $\beta=0$. 
	\end{prop}
	
	Unlike the classical SEM, in the CDC case, either $\alpha$ or $\beta$ must be zero in order for the natural direct effect NDE equal to zero. This is because when $M$ is discrete, $Z_m$ and $M$ are no longer uniquely one-to-one linked. As shown in Web Figure  1,  when only $\gamma$ is zero, $Z_x$ can change $Z_m$ without exerting any change in $M$, and then change in $Z_m$ can cause a change in $Z_y$. This chain of reactions in the latent variable machinery is not uniquely matched with the observed variable machinery due to the discretization, so likely to yield NDE$\neq 0$ under only $\gamma=0$. In other words, discretization may give rise to a certain leakage of causality in such a pathway. According to Proportion~\ref{prop: cdc}, a zero NDE occurs if either $\gamma=\alpha=0$ or $\gamma=\beta=0$. 
	
	\begin{prop}\label{prop: ccd}
		Suppose that $X$ and $M$ given $\bfW$ are normally distributed with marginal means $\mu_x = \bfW_1^\top \bfbeta_x, \mu_m = \bfW^\top \bfbeta_m$ and marginal variances $\sigma_x^2, \sigma_m^2$, and that binary $Y$ given $\bfW$ is Bernoulli distributed with the probability of success satisfying a logistic model of the form, $\mbox{expit}(\bfW^\top \bfbeta_y)$.  Then the mediation causal estimands a change of $X$ from $x_0$ to $x_1$ are given by 
		\begin{align}
			\nde(x_0,x_1; \bfw)&=\int_{-\infty}^{\infty}(\Phi_1(z_{x_0})-\Phi_1(z_{x_1}))\pi(Z_m^*|Z_x=z_{x_0})dZ_m^*,\nonumber\\
			\nie(x_0,x_1; \bfw)&=\int_{-\infty}^{\infty}\Phi_1(z_{x_1})\pi(Z_m^*|Z_x=z_{x_0})dZ_m^*-\int_{-\infty}^{\infty}\Phi_1(z_{x_1})\pi(Z_m^*|Z_x=z_{x_1})dZ_m^*,\nonumber
		\end{align}
		where 
		$\psi(z_{x_a})=\Phi(\tau_yl_y(\bfw)-\gamma z_{x_a}-\tau_m\beta z_m^*)$, $l_y(\bfw)=\Phi^{-1}(F_y(0))=-\Phi^{-1}\left(\mbox{expit}(\bfw^{\top}\bfbeta_y)\right)$. 
		Thus, $\nde(x_0,x_1; \bfw)=0$ if and only if $\gamma=0$, and $\nie(x_0,x_1; \bfw)=0$ if $\alpha=0$ or $\beta=0$. 
	\end{prop}
	Since neither NDE nor NIE above has a closed-form, we resort to the Monte Carlo method to evaluate the integrals by three steps: (i) draw independent copies of $Z^*_m \sim N\left(\frac{\alpha}{\sqrt{\alpha^2+1}} z_{x_b}, \frac{1}{\alpha^2+1}\right)$; (ii) plug each $Z_m^*$ into $\psi(z_{x_a})$; (iii) average over all copies of $\psi(z_{x_a})$.
	
	Conditional $\nde(x_0,x_1; \bfw)$ and $\nie(x_0,x_1; \bfw)$ may be interpreted as risk differences based on odds ratio in the CCD case with binary $Y$. According to  \cite{vanderweele2010odds}, we calculate the odds ratios for the NDE and NIE as follows:
	\begin{align}
		OR^{\nde}(x_0,x_1;\bfw)=\frac{A_{x_0,x_0}/(1-A_{x_0,x_0})}{A_{x_1,x_0}/(1-A_{x_1,x_0})}, ~
		OR^{\nie}(x_0,x_1;\bfw)=\frac{A_{x_1,x_0}/(1-A_{x_1,x_0})}{A_{x_1,x_1}/(1-A_{x_1,x_1})},\label{eq: OR}
	\end{align}
	where $A_{x_a,x_b}=\int_{-\infty}^{\infty}\Phi_1(z_{x_a})\pi(Z_m^*|Z_x=z_{x_b})dZ_m^*$ for $a=0,1$ and $b=0,1.$
	
	\section{Simulation Studies} \label{simul}
	\noindent We conduct  three simulation studies to evaluate the performance of the IFM approach in the proposed GSEM. The first one assesses  bias, mean squared error (MSE) and coverage probability (CP) of confident interval for causal estimands (NED and NIE) as well as model parameters ($\alpha, \beta,\gamma$). The second study compares bias, MSE and CP with a popular existing method \citep{tingley2014mediation} via its R package \texttt{mediation}.  Both experiments were performed under three settings CCC,  CDC and CCD discussed in Section \ref{three eg}. The third study focuses on the CCD case in which we compare the odds ratios calculated from the GSEM and an approximation approach proposed by \cite{vanderweele2010odds}. 
	
	\subsection{Assessment of GSEM}\label{simul: accuracy}
	\noindent Under each of the three settings, we run 1,000 rounds of simulations and calculate CP of a confidence interval by 500 bootstrap samples from both parametric and nonparametric resampling methods. In all settings,  the sample size $n$ varies over 200, 500, and 1000. One GSEM model consists of marginal GLMs and a Gaussian copula with the acyclicity topology under the latent variable representation. For the vectors of covariates $\bfW_1$ and $\bfW$, their first column is set to 1 for intercept and the remaining variables follow multivariate normal distributions with mean zero and compound symmetry correlation $\rho=0.2$ and equal standard deviation $\sigma=0.3$. Set $\bfbeta_x=(0.5,0.2,0.2)^{\top}$, $\bfbeta_m=(0.8,0.3,0.3,0.4)^{\top}$, $\bfbeta_y=(-0.2,0.4,-0.2,0.7)^{\top}$, and $\sigma_x=\sigma_m=\sigma_y=0.3$. The DAG parameters $(\alpha,\gamma,\beta)$ equal to $ (0.20, 0.10, 0.20)$ for CCC, $(0.15, 0.10, 0.75)$ for CDC and $(0.70, 0.10, 0.18)$ for CCD. After generating $Z_x$,  $Z_m^*$, and $Z_y^*$ from the latent model \eqref{eq: general_model}, we generate exposure $X=\bfW_1^{\top}\bfbeta_x+\sigma_x Z_x$, followed by the generation of mediator $M$ and outcome $Y$ as follows. CCC: $M=\bfW^{\top}\bfbeta_m+\sigma_m Z_m^*$, $Y=\bfW^{\top}\bfbeta_y+\sigma_y Z_y^*$; CDC: $M=I\{\Phi(Z_m^*)>\frac{1}{\exp(\bfW^{\top}\bfbeta_m)+1}\}$, $Y=\bfW^{\top}\bfbeta_y+\sigma_y Z_y^*$; and CCD: $M=\bfW^{\top}\bfbeta_m+\sigma_m Z_m^*$, $Y=I\{\Phi(Z_y^*)>\frac{1}{\exp(\bfW\bfbeta_y)+1}\}$.
	
	\begin{center}
		\vskip-0.5cm
		{ [Table 1 about here.]}
		\vskip-0.7cm
	\end{center}
	\begin{table}[htbp]
		\caption{True value, bias, MSE, 95\% coverage by parametric bootstrap (PB), and non-parametric bootstrap (NB) for NDE, NIE, $\alpha$, $\gamma$ and $\beta$  under three settings. The sample size varies over 200, 500 and 1,000, with 1,000 data replicates for each sample size. The confidence interval is obtained by 500 bootstrap replicates. \label{tab: bias}}
		\par
		\vskip .2cm
		\footnotesize
		\centerline{\tabcolsep=3truept\begin{tabular}{clc|ccc|ccc|ccc|ccc}
				\hline
				&&& \multicolumn{3}{c|}{Bias ($\times 10^{-3}$)} & \multicolumn{3}{c|}{MSE ($\times 10^{-3}$)} & \multicolumn{3}{c|}{Coverage (PB, \%)}& \multicolumn{3}{c}{Coverage (NB, \%)} \\[5pt]
				Setting&&True&200&500&1000&200&500&1000&200&500&1000&200&500&1000\\
				\hline
				&NDE& 0.10&3.77& 1.76 &-0.40& 4.99 &1.97  &0.90 &94.2 &95.0 &95.9 &94.6&95.1&96.0 \\
				&NIE &0.04 &0.88&0.79  &0.71& 0.45 &0.16 &0.08 &94.4 &94.6&95.0&93.7 &94.5&94.6\\ 
				CCC&$\alpha$&0.20 &5.29&2.34  &0.54& 5.72 &2.13 &0.98 &93.1 &94.7&95.6&93.4 &94.4&95.1\\
				&$\gamma$&0.10 &4.93&2.05 &-0.33& 5.41 &2.11  &0.98 &94.0&95.1&95.9 &94.5&95.3&96.2 \\
				&$\beta$& 0.20&2.66&2.39  &3.25&5.75  &2.06  &1.11  &93.5 &94.0&93.8&93.7 &94.5&93.8 \\
				\hline
				&NDE&0.11 &0.98& 0.97 &-0.32& 3.96 &1.61  & 0.71&94.6 &95.2&95.7 &95.3&95.4&96.0 \\
				&NIE & 0.05&3.53&1.53  &0.46&1.46  & 0.55 &0.24 &93.3 &93.4&94.3&93.7&93.0&94.4\\ 
				CDC&$\alpha$&0.15 &8.31&2.59  &0.67&10.23  & 3.84 &1.68 &93.3 &94.1&95.0&93.8&93.3&94.9 \\
				&$\gamma$&0.10 &0.69&0.96  &-0.52&8.01  & 3.15 &1.32 &94.7&94.1&95.2&93.9&94.2 &95.4 \\
				&$\beta$ &0.75&22.84& 13.96 &4.04&20.07 & 6.74 &3.43 &92.8 &94.7&94.4&93.6&95.1&94.5 \\
				\hline
				&NDE&0.13&5.40&0.80  &0.59&21.03  &8.01 &4.00 &93.7&94.2 &94.6& 94.1&94.1&94.2 \\
				&NIE &0.14 &3.70&1.64  &1.25&5.45 & 2.17 &1.05 &94.7&93.6 &94.4&94.6 &94.4&94.5 \\ 
				CCD&$\alpha$& 0.70 &9.22&5.64  &3.54& 7.56&3.09  & 1.65&94.7& 94.8&93.6&93.3 &94.5&94.0  \\
				&$\gamma$ &0.10 &5.36&0.89  &0.67&14.15  &5.21  &2.57 &93.8&94.0 &94.8&94.4 &93.9&94.4 \\
				&$\beta$ & 0.18&7.24&2.53  &1.92&9.33  &3.48 &1.71 &93.8&93.7 &94.2&93.5 &94.6&94.5 \\
				\hline
		\end{tabular}}
	\end{table}
	
	The simulation results are summarized in Table \ref{tab: bias}.  Under the three settings, the magnitude of average bias for parameter estimation and causal estimands decrease as the sample size increases. The MSE for $n=1000$ is at most one fifth of that for $n=200$ in each setting. The 95\% confidence intervals have shown proper coverage rates for all $n=200, 500, 1000$  close to the 95\% nominal level using either parametric or non-parametric bootstrap. The computation costs in the simulation studies are between 30 to 60 minutes for a dataset with sample size $n=500$ under the three different settings. 
	
	\subsection{Comparison to QBMC}
	\noindent In the second simulation study, we compare the coverage probability and MSE between our proposed GSEM method and a popular R package \texttt{mediation} \citep{tingley2014mediation,imai2010general}, a quasi-Bayesion Monte Carlo simulation-based method (in short, QBMC).  The comparison is conducted under two scenarios: one under datasets being generated from the  GSEM, and the other from the QBMC. We focus on assessing their performance in terms of estimation bias and CP for NDE and NIE equal to zero (null), and not zero (non-null).  
	
	The GSEM model is specified the same as in Section \ref{simul: accuracy}. Let $\bfbeta_x=(0.5,0.2,0.2)$, $\bfbeta_m=(0.8,0.3,0.3,0.4)$, $\bfbeta_y=(-0.2,0.4,-0.2,0.7)$, and $\sigma_x=\sigma_m=\sigma_y=0.3$. For the null effects, $(\alpha,\gamma,\beta)$ takes the following values: $(0.30, 0, 0)^{\top}$, $(0.15, 0, 0)^{\top}$ and $(0.70, 0, 0)^{\top}$.  According to the propositions discussed in Section \ref{three eg}, the conditional NDE and NIE are both zero in these three settings. For the non-zero causal effects, $(\alpha,\gamma,\beta)$ takes values $(0.20, 0.10, 0.20)$, $(0.15, 0.10, 0.75)$ and $(0.70, 0.10, 0.18)$. 
	
	The QBMC model is set up as follows:  $X \sim N(\bfW_1^{\top}\bfbeta_x,\sigma_X)$ and $M$ and $Y$ are generated as follows: CCC:
	$  M=\beta_{xm}X+\bfW^{\top}\bfbeta_m+\epsilon_M$,
	$Y=\beta_{xy}X+\beta_{my}M+\bfW^{\top}\bfbeta_y+\epsilon_Y
	$;  CDC: $\mbox{logit}P(M=1)=\beta_{xm}X+\bfW^{\top}\bfbeta_m$, $Y=\beta_{xy}X+\beta_{my}M+\bfW^{\top}\bfbeta_y+\epsilon_Y$; and CCD: $M=\beta_{xm}X+\bfW^{\top}\bfbeta_m+\epsilon_M$,
	$\mbox{logit}P(Y=1)=\beta_{xy}X+\beta_{my}M+\bfW^{\top}\bfbeta_y$, where $\sigma_x=\sigma_m=\sigma_y=0.3$, $\bfbeta_x=(0.5,0.2,0.2)$, $\bfbeta_m=(0.5,0.2,0.2,0.4)$, $\bfbeta_y=(-0.8,0.2,-0.5,0.4)$. Moreover, $(\beta_{xm},\beta_{xy}, \beta_{my})
	= (0.30, 0, 0), (0.30, 0, 0), (0.30, 0, 0)$ for the scenarios of null mediation effects, and $(\beta_{xm},\beta_{xy}, \beta_{my})
	= (0.20, 0.10, 0.20), (0.40, 0.10, 0.50), (0.70, 0.50, 0.70)$ for the scenarios of non-zero mediation effects.

	Table \ref{tab: coverage comparison} summarizes the comparison results between GSEM and QBMC for their empirical CP, average bias and MSE. When the data are generated from the GSEM, the coverage rates by the IFM are closer to the nominal 95\% level in most cases. For average bias and MSE, the IFM outperformed QBMC in some cases, such as CCD where the GSEM consistently provides smaller bias and MSE for NIE regardless of null or non-null mediation effects. In the CDC case, the IFM provides smaller bias  for NIE and NDE compared to QBMC when sample size $n=500$ under the non-null settings. For the other scenarios, the results between the two methods are comparable. 
	
	When the data are generated  from QBMC, in the cases of null mediation effects two methods are comparable. In the cases of non-null mediation effects, since the QBMC method does not have formulas to calculate the true effects, we estimated the true effects from a simulated dataset of $n=100,000$, under which the resulting estimates of NDE and NIE from the R package are treated as the true effect sizes. Thus,  the comparisons under the nonzero mediation effect settings need to be interpreted cautiously.  
	\begin{center}
		{ [Table 2 about here.]}
		\vskip-2cm
	\end{center}
	\begin{table}[htbp]
		\caption{Comparisons between IFM and QBMC for coverage rate via the parametric bootstrap, average bias and MSE summarized over 1000 rounds of simulations, where 500 bootstrap samples are used in the construction of confidence interval. 
			The sample size $n=200, 500$.  \label{tab: coverage comparison}}
		\par
		\vskip .2cm
		\footnotesize
		\centerline{\tabcolsep=3truept\begin{tabular}{cclcl|cccccc}
				\hline
				&&&&& \multicolumn{2}{c}{Coverage (\%)} & \multicolumn{2}{c}{Bias ($\times 10^{-3}$)} &\multicolumn{2}{c}{MSE ($\times 10^{-3}$)} \\[5pt]
				Simulation Model&Setting&Effect&True&Method&200&500&200&500&200&500\\
				\hline
				\multirow{24}{*}{GSEM}&\multirow{8}{*}{CCC}&\multirow{2}{*}{NDE}&\multirow{2}{*}{0}&IFM &95.0 &94.6& -1.70&  3.62& 5.82& 2.31 \\
				&&&&QBMC  &95.4& 94.4&-1.68&  3.59& 5.67& 2.27\\
				
				&&\multirow{2}{*}{NIE}&\multirow{2}{*}{0}&IFM&92.6& 94.3&-0.30&-0.37&0.58& 0.20\\ &&&&QBMC&93.2&94.0 &-0.26& -0.36& 0.54& 0.20\\
				&&\multirow{2}{*}{NDE}&\multirow{2}{*}{0.10}& IFM & 95.3 &94.8 &-0.56&  3.89& 5.13& 2.06\\
				&&&&QBMC&95.4&94.0 &-1.74& 3.37& 5.03& 2.03\\
				&&\multirow{2}{*}{NIE}&\multirow{2}{*}{0.04}&IFM &94.0& 94.6&1.28& -0.12& 0.47& 0.16\\
				&& && QBMC &94.0 &94.7 &-0.03& -0.65& 0.44& 0.16\\
				\cline{2-11}
				&\multirow{8}{*}{CDC}&\multirow{2}{*}{NDE}&\multirow{2}{*}{0}&IFM & 95.1&94.0&-1.97&3.40& 5.24& 2.12  \\
				&&&&QBMC  &95.1&93.8 &-1.96& 3.35& 5.07& 2.09\\
				&&\multirow{2}{*}{NIE}&\multirow{2}{*}{0}&IFM &97.8& 96.0&-0.02& -0.15&0.12& 0.04\\
				&&&&QBMC&98.7& 96.9&0.02&-0.12& 0.10& 0.03\\
				&&\multirow{2}{*}{NDE}&\multirow{2}{*}{0.11}& IFM & 93.9 &94.2 &-1.87& 2.23& 4.18& 1.64 \\
				&&  &&QBMC&94.2& 94.3&0.51&  4.30& 4.14& 1.66\\
				&& \multirow{2}{*}{NIE}&\multirow{2}{*}{0.05}&IFM &93.9& 95.0&  3.01&0.05 &1.37& 0.49 \\
				& & && QBMC &96.0 &95.5&-2.77& -3.33& 1.08& 0.42\\
				\cline{2-11}
				&\multirow{8}{*}{CCD}&\multirow{2}{*}{NDE}&\multirow{2}{*}{0}&IFM &96.0&  95.4&  9.14&  1.03& 19.94& 7.70 \\
				&&&&QBMC  &96.6&  95.9&  9.68&  1.86& 17.23& 7.13\\
				&&\multirow{2}{*}{NIE}&\multirow{2}{*}{0}&IFM &94.0 & 94.3& -6.63& -0.12&  4.82& 1.82  \\
				&&&&QBMC&95.4 & 95.0& -8.28& -0.83&  5.82& 2.44 \\
				&&\multirow{2}{*}{NDE}&\multirow{2}{*}{0.13}& IFM &   95.4&  94.6&  6.42&  1.63& 19.35& 7.84 \\
				&&&&QBMC& 95.5&  94.9&  1.39& -1.96& 16.74& 7.20\\
				& &\multirow{2}{*}{NIE}&\multirow{2}{*}{0.14}&IFM &93.6&  95.0& -0.90&  0.87&  5.51& 2.10 \\
				& & && QBMC & 95.3  &93.8&  7.09& 17.24& 5.83& 2.73 \\
				\hline
				\multirow{24}{*}{SEM}&\multirow{8}{*}{CCC}&\multirow{2}{*}{NDE}&\multirow{2}{*}{0}&IFM & 95.0&  94.6& -1.70& 3.62& 5.82& 2.31   \\
				&&&&QBMC  &95.5&94.5& -1.68& 3.59& 5.67& 2.27 \\
				&&\multirow{2}{*}{NIE}&\multirow{2}{*}{0}&IFM&92.6&  94.3& -0.30& -0.37& 0.58&0.20
				\\&&&&QBMC&93.4 & 93.9 &-0.26& -0.36& 0.54& 0.20\\
				&&\multirow{2}{*}{NDE}&\multirow{2}{*}{0.10}& IFM & 95.3 &94.0 & 3.80&  8.38& 5.45& 2.24 \\
				& & &&QBMC&95.1& 94.6 & 2.59&  7.85& 5.33& 2.20  \\
				& &\multirow{2}{*}{NIE}&\multirow{2}{*}{0.04}&IFM & 93.6&94.2 &-0.26& -1.70& 0.49 &0.17\\
				&&& &QBMC &93.7& 93.4 &-1.61& -2.25& 0.47& 0.17\\
				\cline{2-11}
				&\multirow{8}{*}{CDC}&\multirow{2}{*}{NDE}&\multirow{2}{*}{0}&IFM & 94.0& 93.9 &-4.06& -0.98& 5.16& 2.06   \\
				&&&&QBMC  &95.2& 94.1& -4.14& -0.96& 5.02& 2.03   \\
				&&\multirow{2}{*}{NIE}&\multirow{2}{*}{0}&IFM&99.5& 99.3 & 0.34& -0.02& 0.05& 0.01 \\
				&&&&QBMC&99.9 &99.6& 0.34& -0.02& 0.04& 0.01\\
				&&\multirow{2}{*}{NDE}&\multirow{2}{*}{0.09}& IFM & 93.6 &94.0& 5.24& 7.49 &5.28 &2.17\\
				&&&& QBMC &95.7& 93.5& 1.95& 5.13 &5.01& 2.05\\
				& &\multirow{2}{*}{NIE}&\multirow{2}{*}{0.04}& IFM &93.5& 93.6& 5.41& 3.76& 3.32& 1.23   \\
				&&&&QBMC& 95.7& 96.7 &5.41 &4.83& 3.31& 1.24  \\
				\cline{2-11}
				&\multirow{8}{*}{CCD}&\multirow{2}{*}{NDE}&\multirow{2}{*}{0}&IFM &94.3& 94.2&  1.08&  3.93& 13.72& 5.42   \\
				&&&&QBMC  & 95.6 &94.5&  1.91&  4.10& 12.72& 5.21   \\
				&&\multirow{2}{*}{NIE}&\multirow{2}{*}{0}& IFM &93.7& 95.1& -2.71& -0.79&  1.17& 0.40 \\
				&&&&QBMC&95.2& 94.8& -3.41& -1.21 & 1.10& 0.42  \\
				&&\multirow{2}{*}{NDE}&\multirow{2}{*}{0.12}& IFM & 94.3& 95.3&  11.24&   7.20 &21.25 &8.26 \\
				&&&&QBMC&95.0 &95.5&   5.36   &3.70& 18.25& 7.63 \\
				& &\multirow{2}{*}{NIE}&\multirow{2}{*}{0.12}& IFM f&94.2& 93.2& -21.14& -18.14 & 5.46& 2.19 \\
				&& && QBMC &95.7& 96.5& -12.30&  -4.60 & 5.75& 2.31\\
				\hline
		\end{tabular}}
	\end{table}
	
	\subsection{Odds Ratio comparison for Binary Outcome}
	\noindent The third simulation concerns the case of CCD, where the NDE and NIE on the odds ratio scale are compared between the GSEM version \eqref{eq: OR} and an approximation approach proposed by \cite{vanderweele2010odds} (termed as ``VV" method for convenience).  We consider two scenarios: a rare prevalence with around 4\% of outcomes being ``1" and an abundant prevalence with around 45\% of outcomes being ``1". 
	
	For both scenarios, data are generated from the respective GSEMs with the same $\bfbeta_x$ and $\bfbeta_m$ as well as the same covariates $\bfW$ and $\bfW_1$ and variance parameters as those in the previous two simulation studies. In addition,  set $(\alpha, \beta, \gamma) = (0.70, 0.18, 0.10)$. For the rare prevalence,  $\bfbeta_y=(-0.2,0.4,-0.2,0.7)$, while for the case of abundant prevalence, $\bfbeta_y=(-3.0,0.4,-0.2,0.7)$.  See the results in Table~\ref{tab: OR} for the odds ratio of NDE and NIE. 
	\begin{center}
		
		{ [Table 3 about here.]}
		\vskip-2cm
	\end{center}
	\begin{table}[htbp]
		\caption{True value, mean bias, MSE  comparison of $OR^{NDE}$ and  $OR^{NIE}$ for methods GSEM and ``VV". Data are  generated from GSEM, and sample size varies over 500, 1,000, and 2,000 with 1000 data replicates for each sample size.  \label{tab: OR}}
		\par
		\vskip .2cm
		\footnotesize
		\centerline{\tabcolsep=3truept\begin{tabular}{clcc|ccc|ccc}
				\hline
				&&&& \multicolumn{3}{|c|}{Bias } & \multicolumn{3}{c}{MSE} \\[5pt]
				Setting&Odds Ratio&True&Method&500&1000&2000&500&1000&2000\\
				\hline
				\multirow{ 4}{*}{Abundant}&\multirow{2}{*}{$OR^{NDE}$}&\multirow{2}{*}{1.704}& GSEM&0.116   & 0.098&    0.022&   0.573    &0.250    &0.091 \\
				&& &VV&  0.134 &   0.114&    0.035&   0.600&0.264&    0.095\\ 
				&\multirow{2}{*}{$OR^{NIE}$}&\multirow{2}{*}{1.762} &GSEM& 0.070 &   0.026&    0.017&   0.136 &   0.056&    0.027\\
				&& &VV &0.315   & 0.251&    0.238 &  0.349&    0.166&    0.105\\
				\hline
				\multirow{ 4}{*}{ Rare}&\multirow{2}{*}{$OR^{NDE}$}&\multirow{2}{*}{2.169}& GSEM& 1.135 &   0.662&    0.290&  22.062&    5.109&    1.629\\
				&&& VV  &0.618  &  0.350&    0.078&  10.039&    2.870&    1.055\\ 
				&\multirow{2}{*}{$OR^{NIE}$}& \multirow{2}{*}{2.113}&GSEM & 0.502  &  0.157&    0.087&   2.662&    0.660&    0.280 \\
				&&& VV &0.830  &  0.490&    0.421&   3.610&    1.194&    0.591\\
				
				\hline
		\end{tabular}}
	\end{table}
	
	For the case of abundant events, the VV method has higher average bias and MSE than the IFM under the GSEM, demonstrating a clear out-performance of IFM over VV.   For the case of rare events, the VV method has smaller bias and MSE for the odds ratio of NDE, but larger bias and MSE for the odds ratio of NIE. One explanation for the under-performance of IFM on NDE is likely to be unstable numerical operations in the marginal logistic regression with a very unbalanced rate of events.   Nevertheless, IFM in the GSEM still provides better performance compared to the VV method on the odds ratio of NIE.

		
				
				
	
	\section{Data Application}\label{data analysis}
	\noindent We now analyze the motivating data from the ELEMENT cohort study; see the related detail in Section \ref{real data}.  In this analysis, we focus on the prenatal exposure to phthalates during the second trimester (T2).  We hypothesize the influence of such T2 exposure to phthalates on children's cardiometabolic outcome MetZ score in peripuberty may be mediated by the timing of  children reaching their infancy BMI peaks on time ($M=0$) or delayed ($M=1$). This mediator is deemed as an important early-life biomarker to characterize growth tempo. 
	
	Exposure $X$ includes three maternal urinary phthalate  concentrations of MEHHP, MEOHP and MIBP measured at their T2 pregnancy.  The raw phthalate concentrations are right skewed; after log-transforming $X$, we checked that the resulting $\log(X)$ appears normally distributed. Outcome $Y$ includes MetS z-score, which appears normally distributed. The mediator of infant growth marker is binary, where 27.8\% of children have delayed infancy BMI peak time. See the details of the vectors of confounders $\bfW_1$ and $\bfW_2$ in Section~\ref{real data}. 
	
\begin{figure}
		\centerline{
			\begin{tabular}{cc}		
				\psfig{figure=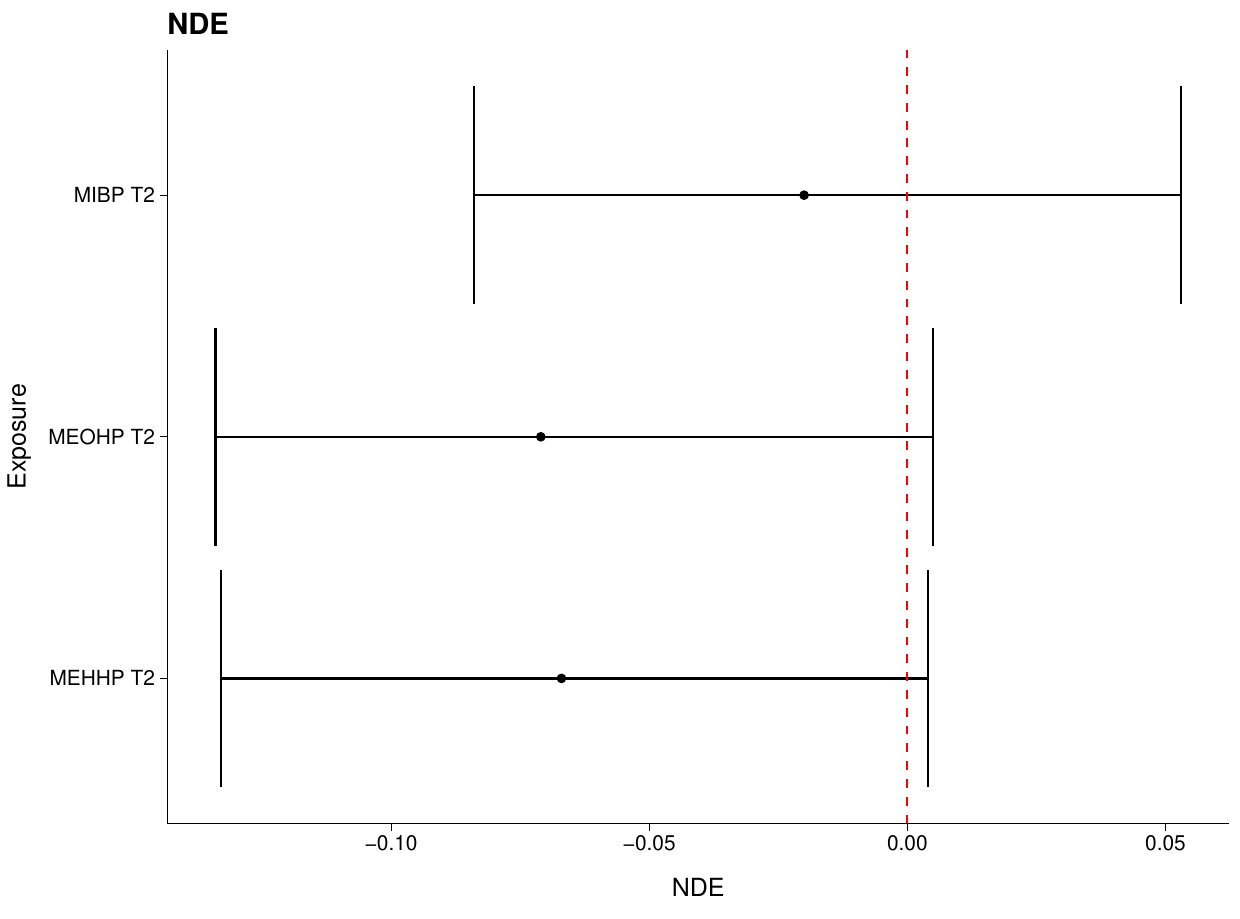,width=3in,angle=0} & \psfig{figure=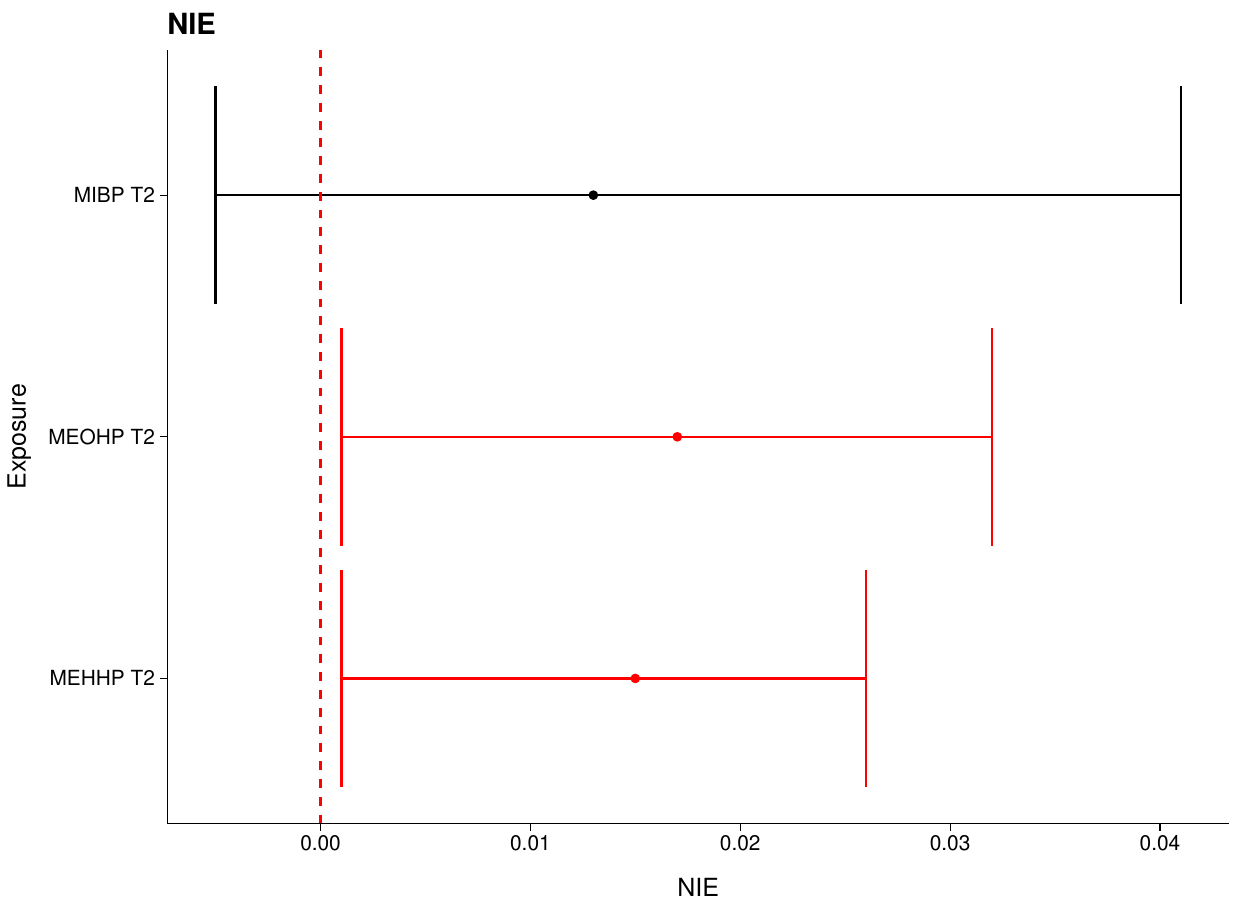,width=3in,angle=0}   \\
				(a) NDE  &  (b) NIE      \\
				\multicolumn{2}{c}{\psfig{figure=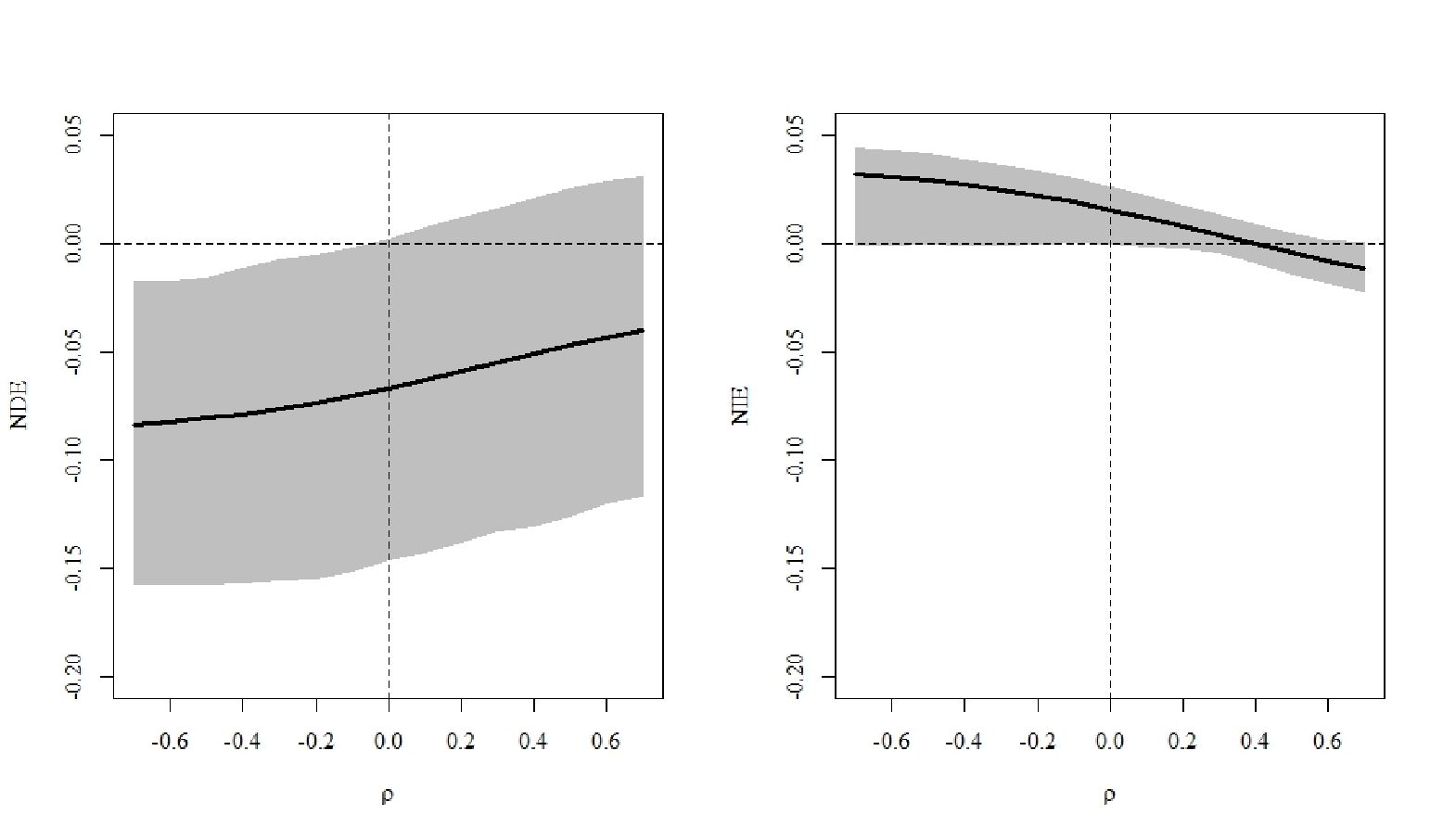,width=6in,angle=0}} 
				\\
				\multicolumn{2}{c}{(c) Sensitivity of violation on independence}  \\ 
				& \\
			\end{tabular}
		}
		\caption{Panels (a)-(b) show the NDE and NIE estimates with 95\% CIs from the mediation analysis of the ELEMENT cohort data using the CDC version of the GSEM. Panel (c) presents NDE and NIE estimates with 95\% CIs against $\rho$ for the sensitivity analysis of the mediation pathway under exposure to phthalate MEHHP. 
		}
		\label{fig: real data}
	\end{figure}

	We apply the CDC version of the GSEM to perform the mediation analysis in order to answer the above scientific hypothesis, where $\bfW_1$ is included in the marginal GLM for continuous exposure $X$, while $\{\bfW_1, \bfW_2\}$ is included in the respective GLMs for binary mediator $M$ and continuous $Y$. We estimate the model parameters and causal estimands by the IFM method and further obtain 95\% CIs for NDE and NIE using the parametric bootstrap. From the results of the confidence intervals summarized in Figure \ref{fig: real data} (a)--(b), two positive NIEs are detected at the significance level $0.05$. The first concerns the mediation pathway under exposure to phthalate MEHHP, with an estimated NIE of 0.015 and 95\% CI (0.001, 0.026). 
	The other positive NIE is found significant at the significant level 0.05 in the mediation pathway under the second trimester exposure to MEOHP, with an estimate of 0.017  and 95\% CI (0.001, 0.032). 
	
	Before finalizing our answer to the hypothesis above, we evaluate the sensitivity of these causal effect estimations to some mild violations of the sequential ignorability assumption. Here we focus on exposure MEHHP in this sensitivity analysis.  To do so, we consider an equivalent model representation of \eqref{eq: general_model}:
	$
	Z_x \sim N(0, 1)$; 
	$Z_m = \alpha Z_x + \epsilon_m ~\mbox{with}~ \epsilon_m \sim N(0, 1)$; and
	$Z_y = \gamma Z_x + \beta Z_m + \epsilon_y~\mbox{with}~\epsilon_y \sim N(0,1)$, 
	where $\epsilon_m$ and $\epsilon_y$ are independent under the model assumption of \eqref{eq: general_model}. To relax this independence assumption,  we now allow $(\epsilon_m, \epsilon_y) \sim BVN(0,0,\sigma_m^2,\sigma_y^2,\rho)$ with correlation $\rho$, under which we estimate both NDE and NIE over a range of $\rho$ values from $-0.7$ to $0.7$ with an increment $0.1$. Note that the case of $\rho = 0$ corresponds to model \eqref{eq: general_model} assumed in the above data analysis. Figure \ref{fig: real data} (c) presents both NDE and NIE estimates against $\rho$ along with their pointwise 95\% confidence intervals. This sensitivity analysis suggests that the GSEM estimates and 95\% CIs are robust to mild violations from model \eqref{eq: general_model} with small deviations of $\rho$ values from the 0. When $\rho$ increases, the estimates of NIE in the mediation pathway under prenatal exposure to MEHHP appear to be firmly positive, steadily staying above the zero horizontal line until the $\rho$ reaches 0.4 or larger. Thus, the above findings from the mediation analysis are trustworthy.

	\begin{center}
		\vskip-0.6cm
		{ [Figure 3 about here.]}
	\end{center}

	\section{Concluding Remarks}\label{concluding}
	\noindent Despite much effort in the literature on the extension of the classical structural equation modeling (SEM) approach to deal with non-normal data of mixed types, there remains little development for a unified framework to address this technical challenge systematically. The proposed GSEM framework has delivered one desirable solution whose pros and cons have been thoroughly investigated in this paper. Being different from all existing works, our proposed GSEM implements one important feature by explicitly embedding the DAG acyclicity topology into the modeling structures, which presents a natural extension of the classical SEM and, more importantly, ensures the hypothesized causal structure of mediation pathway for better interpretability.  In addition, the proposed hierarchical construct of the GSEM allows practitioners to add high-dimensional vectors of confounders in the marginal GLMs for exposure, mediator and outcome, for which existing software packages such as the R package \texttt{glmnet} can readily handle. 
	
	We illustrated three versions of the GSEMs of practical importance where we discovered the issue of potential mediation leakage when categorical variables are involved in the DAG. Such leakage phenomenon requires attention and caution when performing mediation analyses in practice. We also demonstrated and confirmed the asymptotic efficiency of the method of inference function for marginals (IFM) through extensive simulation studies. This profile likelihood estimation has been shown its capacity to accurately make statistical inference on NDE and NIE in the GSEM framework.
	
	
	The current procedure focuses mainly on one-dimensional mediator, and an extension to 
	multi-mediators of mixed types is an important future work. Despite some methods developed for continuous multi-mediators \citep{he2023AdaptiveBootstrapTests,hao2024SimultaneousLikelihoodTest}, there lacks suitable toolboxes for multi-mediators of mixed types. The flexibility of the copula dependent model in the construction of the GSEM allowing different types of marginal distributions seems to be extendable to accommodate mixed types of multiple mediators.  The proposed GSEM may be also extended to deal with survival outcomes using the Clayton copula \citep{oakes1986SemiparametricInferenceModel}  and to develop quantile SEM framework in the future work. 

	
	\backmatter
	\vspace*{-8pt}
	
	\section*{Acknowledgements}
	This research work was supported by grant U24ES028502 and R01ES033656. The authors thank Dr. Karen Peterson for constructive discussions on the data application.
	\vspace*{-8pt}
	
	
	\section*{Supporting Information}
	{
		Technical derivations and additional tables mentioned in the main text are available as web appendices on the Biometrics website of Wiley Online Library.
	}
	\vspace*{-8pt}
	
	
	\bibliographystyle{biom}
	\bibliography{ref}
	
	\label{lastpage}
	
\end{document}